\documentclass[seceq,preprint]{ptptex}
\usepackage{graphicx}

\usepackage{wrapft}

\newcommand{\be}{\begin{equation}}
\newcommand{\ee}{\end{equation}}

\newcommand{\bea}{\begin{eqnarray}}
\newcommand{\eea}{\end{eqnarray}}

\newcommand{\Tr}{{\rm Tr}}

\pubinfo{Vol.~???, No.~?, Month YYYY}
\preprintnumber[4.0cm]{
IFUP--TH/2010--33 \\ Saclay--IPhT--t10/152 \\ October 2010}

\title{
High-energy hadron-hadron (dipole-dipole) scattering on the lattice
}

\author{
Enrico \textsc{Meggiolaro}$^{1,}$\footnote{Speaker at the Symposium.
E--mail: enrico.meggiolaro@df.unipi.it}
and Matteo \textsc{Giordano}$^{1,2,}$\footnote{Supported by a grant of the
``Fondazione Angelo Della Riccia'' (Florence, Italy).}
}

\inst{
$^1$Dipartimento di Fisica, Universit\`a di Pisa, and INFN, Sezione di Pisa,\\
Largo Pontecorvo 3, I-56127 Pisa, Italy\\
$^2$ Institut de Physique Th\'eorique CEA-Saclay,\\ F-91191 Gif-sur-Yvette
Cedex, France
}

\recdate{
October ??, 2010
}

\abst{
We will discuss how the problem of high--energy hadron--hadron
(dipole--dipole) scattering at low momentum transfer can be approached from the
point of view of lattice QCD, by means of Monte Carlo numerical simulations.
}

\begin{document}
\maketitle

\section{Introduction}

The problem of predicting from first principles total cross sections
at high energy is one of the oldest open problems of hadronic physics
(see, for example, Ref.~\cite{pomeron-book} and references therein).
As QCD is believed to be the fundamental theory of strong
interactions, it should predict the correct asymptotic behaviour:
nevertheless, a satisfactory explanation is still lacking.
The problem of total cross sections is part of the more
general problem of high--energy elastic scattering at low
transferred momentum, the so--called {\it soft high--energy
scattering}. As soft high--energy processes possess two different
energy scales, the total center--of--mass energy squared 
$s$ and the transferred momentum squared $t$,
smaller than the typical energy scale of strong
interactions ($|t| \lesssim 1~ {\rm GeV}^2 \ll s$), we cannot fully rely on
perturbation theory. A genuine nonperturbative approach in the framework of
QCD has been proposed by Nachtmann in Ref.~\cite{Nachtmann91} and further
developed in Refs.:\cite{DFK,Nachtmann97,BN,Dosch,LLCM1} using a functional
integral approach, high--energy hadron--hadron elastic scattering
amplitudes are shown to be governed by the correlation function of
certain Wilson loops defined in Minkowski space. Moreover, as it has
been shown in Refs.,\cite{Meggiolaro97}\tocite{GM2009} such a correlation
function can be reconstructed by analytic continuation from its
Euclidean counterpart, i.e., the correlation function of two Euclidean
Wilson loops, that can be calculated using the nonperturbative
methods of Euclidean Field Theory.

In Refs.~\cite{GM2008,GM2010} we have investigated this problem 
by means of numerical simulations in Lattice Gauge Theory (LGT).
Although we cannot obtain an analytic expression in this way, nevertheless
this is a first--principle approach that provides (inside the errors)
the true QCD expectation for the relevant correlation function.
In this contribution,
after a quick survey of the nonperturbative approach to soft
high--energy scattering in the case of meson--meson {\it elastic} scattering,
we will present our numerical approach based on LGT, and we will show how
the numerical results can be compared to the existing analytic models.

\section{High--energy meson--meson elastic scattering amplitudes and
Wilson--loop correlators}

We sketch here the nonperturbative approach to soft high--energy scattering
(see Ref.~\cite{GM2008} for a more detailed presentation).
The elastic scattering amplitudes of two mesons (taken for simplicity
with the same mass $m$) in the {\it soft} high--energy regime can be
reconstructed, after folding with the appropriate wave functions, from
the scattering amplitude ${\cal M}_{(dd)}$ of two dipoles of fixed transverse
sizes $\vec{R}_{1\perp}$, $\vec{R}_{2\perp}$, and fixed longitudinal--momentum
fractions $f_1$, $f_2$ of the two quarks in the two dipoles:\cite{DFK}
\be
{\cal M}_{(dd)} (s,t;1,2) 
\equiv -i~2s \displaystyle\int d^2 \vec{z}_\perp
e^{i \vec{q}_\perp \cdot \vec{z}_\perp}
{\cal C}_M(\chi \mathop\sim_{s \to \infty} \log(s/m^2);
\vec{z}_\perp;1,2) ,
\label{scatt-loop}
\ee
where $s \equiv (p_1 + p_2)^2$ and $t = -|\vec{q}_\perp|^2$ ($\vec{q}_\perp$
being the transferred momentum) are the usual Mandelstam variables,
and the arguments ``$1$'' and ``$2$'' stand for ``$\vec{R}_{1\perp}, f_1$'' and
``$\vec{R}_{2\perp}, f_2$'' respectively.
The correlation function ${\cal C}_M$ is defined as the
limit $\displaystyle {\cal C}_M \equiv \lim_{T\to\infty} {\cal G}_M $
of the correlation function of two loops of finite length $2T$,
\be
{\cal G}_M(\chi;T;\vec{z}_\perp;1,2) \equiv
\dfrac{\langle {\cal W}^{(T)}_1 {\cal W}^{(T)}_2 \rangle}{\langle
{\cal W}^{(T)}_1 \rangle \langle {\cal W}^{(T)}_2 \rangle } - 1,
\label{GM}
\ee
where $\langle\ldots\rangle$ are averages in the sense of the QCD
functional integral, and
\be
{\cal W}^{(T)}_{1,2} \equiv
{\dfrac{1}{N_c}} \Tr \left\{ {\cal P} \exp
\left[ -ig \displaystyle\oint_{{\cal C}_{1,2}} A_{\mu}(x) dx^{\mu} \right]
\right\} 
\label{QCDloops}
\ee
are Wilson loops in the fundamental representation of $SU(N_c=3)$; the
paths are made up of the classical trajectories of quarks and
antiquarks, 
\be
{\cal C}_1 : X^{1q[\bar{q}]}_{}(\tau) = z + \dfrac{p_{1}}{m} \tau
+ f^{q[\bar{q}]}_1 R_{1} , \quad
{\cal C}_2 : X^{2q[\bar{q}]}_{}(\tau) = \dfrac{p_{2}}{m} \tau
+ f^{q[\bar{q}]}_2 R_{2},
\label{traj}
\ee
with $\tau\in [-T,T]$, and closed by straight--line paths in the
transverse plane at $\tau=\pm T$ in order to ensure gauge invariance. 
Here
\be 
{p_1}={m}\Big( \cosh \displaystyle\dfrac{\chi}{2},\sinh
\dfrac{\chi}{2},\vec{0}_\perp \Big) ,~~~
{p_2}={m}\Big( \cosh \displaystyle \dfrac{\chi}{2},-\sinh
\dfrac{\chi}{2},\vec{0}_\perp \Big), 
\label{p1p2}
\ee
$\chi$ being the hyperbolic angle formed by the two trajectories, i.e.,
$p_1 \cdot p_2 = m^2 \cosh\chi$.
Moreover, $R_1 = (0,0,\vec{R}_{1\perp})$, $R_2 = (0,0,\vec{R}_{2\perp})$,
$z = (0,0,\vec{z}_\perp)$, and $f^{q}_i = 1-f_i$, $f^{\bar{q}}_i = -f_i$
($i=1,2$), with $f_i \in [0,1]$ the longitudinal--momentum fraction of
quark ``$i$''.\\
The Euclidean counterpart of Eq.~(\ref{GM}) is
\be
{\cal G}_E(\theta;T;\vec{z}_\perp;1,2) \equiv
\dfrac{\langle \widetilde{\cal W}^{(T)}_1 \widetilde{\cal W}^{(T)}_2 \rangle_E}
{\langle \widetilde{\cal W}^{(T)}_1 \rangle_E
\langle \widetilde{\cal W}^{(T)}_2 \rangle_E } - 1,
\label{GE}
\ee
where now $\langle\ldots\rangle_E$ is the average in the sense of the
Euclidean QCD functional integral, and the Euclidean Wilson loops
\be
\widetilde{\cal W}^{(T)}_{1,2} \equiv
{\displaystyle\dfrac{1}{N_c}} \Tr \left\{ {\cal P} \exp
\left[ -ig \displaystyle\oint_{\widetilde{\cal C}_{1,2}}
A_{E\mu}(x_E) dx_{E\mu} \right] \right\}
\label{QCDloopsE}
\ee
are calculated on the following straight--line paths,
\be
\widetilde{\cal C}_1 : X^{1q[\bar{q}]}_{E}(\tau) = z + \frac{p_{1E}}{m} \tau
+ f^{q[\bar{q}]}_1 R_{1E} , \quad
\widetilde{\cal C}_2 : X^{2q[\bar{q}]}_{E}(\tau) = \frac{p_{2E}}{m} \tau
+ f^{q[\bar{q}]}_2 R_{2E},
\label{trajE}
\ee
with $\tau\in [-T,T]$, and closed by straight--line paths in the
transverse plane at $\tau=\pm T$.
The four--vectors $p_{1E}$ and $p_{2E}$ are chosen to be (taking $X_{E4}$ to be
the ``Euclidean time'')
\be
{p_{1E}}={m}\Big( \sin\frac{\theta}{2}, \vec{0}_{\perp},
\cos\frac{\theta}{2} \Big), \quad
{p_{2E}}={m}\Big( -\sin\frac{\theta}{2}, \vec{0}_{\perp},
\cos\frac{\theta}{2} \Big),
\label{p1p2E}
\ee
$\theta$ being the angle formed by the two trajectories, i.e.,
$p_{1E} \cdot p_{2E} = m^2 \cos\theta$.
Moreover, $R_{1E} = (0,\vec{R}_{1\perp},0)$, $R_{2E} = (0,\vec{R}_{2\perp},0)$
and $z_E = (0,\vec{z}_{\perp},0)$ (the transverse vectors are taken to be equal
in the two cases).
Again, we define the correlation function with the IR cutoff removed
as $\displaystyle {\cal C}_E \equiv \lim_{T\to\infty} {\cal G}_E $. 

It has been shown that the correlation functions in the two theories are
connected by the {\it analytic--continuation relations}:
\cite{Meggiolaro97}\tocite{GM2009}
\bea
{\cal G}_M(\chi;T;\vec{z}_\perp;1,2)
&=& \overline{\cal G}_E (-i\chi;iT;\vec{z}_\perp;1,2) ,
\qquad \forall\chi\in \mathbb{R}^+,\nonumber \\
{\cal G}_E(\theta;T;\vec{z}_\perp;1,2)
&=& \overline{\cal G}_M (i\theta;-iT;\vec{z}_\perp;1,2) ,
\qquad \forall\theta\in (0,\pi).
\label{analytic}
\eea
Here we denote with an overbar the analytic extensions of the Euclidean and
Min\-kow\-skian correlation functions, starting from the real intervals
$(0,\pi)$ and $(0,\infty)$ of the respective angular variables, with positive
real $T$ in both cases, into domains of the complex variables $\theta$ 
(resp.~$\chi$) and $T$ in a two--dimensional complex space.
(See Ref.~\cite{GM2009} for a more detailed discussion: in particular,
in Ref.~\cite{GM2009} we have shown, on nonperturbative grounds, that the
required analyticity hypotheses are indeed satisfied, thus obtaining a real
nonperturbative foundation of Eqs. \eqref{analytic}.)\\
Under certain analyticity hypotheses in the $T$ variable, the
following relations are obtained for the correlation functions with
the IR cutoff $T$ removed:\cite{Meggiolaro05,GM2009}
\bea
{\cal C}_M(\chi;\vec{z}_\perp;1,2) &=&
\overline{\cal C}_E(-i\chi;\vec{z}_\perp;1,2) ,
\qquad \forall\chi\in \mathbb{R}^+,\nonumber \\
{\cal C}_E(\theta;\vec{z}_\perp;1,2) &=&
\overline{\cal C}_M(i\theta;\vec{z}_\perp;1,2) ,
\qquad \phantom{-}\forall\theta\in (0,\pi).
\label{analytic_C}
\eea
Finally, we recall the so--called {\it crossing--symmetry
relations}:\cite{GM2006,Meggiolaro07,GM2009}
\bea
&\overline{\mathcal{C}}_M(i\pi-\chi;\vec{z}_{\perp};1,2)
=\mathcal{C}_M(\chi;\vec{z}_{\perp};1,\overline{2}) 
=\mathcal{C}_M(\chi;\vec{z}_{\perp};\overline{1},2) ,
\quad \forall\chi\in \mathbb{R}^+,&
\nonumber \\
& \mathcal{C}_E(\pi-\theta;\vec{z}_{\perp};1,2)
=\mathcal{C}_E(\theta;\vec{z}_{\perp};1,\overline{2}) 
~=\mathcal{C}_E(\theta;\vec{z}_{\perp};\overline{1},2) ,
\quad\forall\theta\in (0,\pi).&
\label{crossing_C}
\eea
Here the arguments ``$\overline{i}$'' stand for ``$-\vec{R}_{i\perp}, 1-f_i$''
($i=1,2$): the exchange ``$1,2$'' $\to$ ``$1,\overline{2}$'', or ``$1,2$'' $\to$
``$\overline{1},2$'', corresponds to the exchange from a loop--loop correlator
to a loop--{\it antiloop} correlator, where an {\it antiloop} is obtained from
a given loop by exchanging the quark and the antiquark trajectories.\\
In the following, we will take for simplicity the longitudinal--momentum
fractions $f_1$, $f_2$ of the two quarks in the two dipoles to be fixed to
$1/2$: as it is explained in the Appendix of Ref.,\cite{GM2010} one can always
reduce to this case without loss of generality.
We will also adopt the notation
${\cal G}_E(\theta;T;\vec{z}_{\perp};\vec{R}_{1\perp},\vec{R}_{2\perp})\equiv
{\cal G}_E(\theta;T;\vec{z}_{\perp};\vec{R}_{1\perp},f_1=\frac{1}{2},
\vec{R}_{2\perp},f_2=\frac{1}{2})$, and similarly for ${\cal C}_E$.

\section{Wilson--loop correlators on the lattice}

The gauge--invariant Wilson--loop correlation function
${\cal G}_E$ is a natural candidate for a lattice computation, but
some care has to be taken due to the explicit breaking of $O(4)$
invariance on a lattice. As straight lines on a lattice can be either parallel
or orthogonal, we are forced to use {\it off--axis} Wilson loops to
cover a significantly large set of angles.\cite{GM2008} To stay as close as
possible to the continuum case, the loop sides are evaluated on the
lattice paths that minimise the distance from the true, 
continuum paths: this can be easily accomplished making use of the well--known
{\it Bresenham algorithm}~\cite{Bres} to find the required
``{\it minimal distance paths}'' corresponding to the sides of the loops.
The relevant Wilson loops
$\widetilde{\cal W}_L(\vec{l}_{\parallel};\vec{r}_{\perp};n)$
are then characterised by the position $n$ of their center and by two
two--dimensional vectors $\vec{l}_{\parallel}$ and $\vec{r}_{\perp}$,
corresponding respectively to the longitudinal and transverse sides of
the loop.

On the lattice we then define the correlator
\be
\label{eq:corr_lat}
{\cal G}_L(\vec{l}_{1\parallel},\vec{l}_{2\parallel};\vec{d}_{\perp};
\vec{r}_{1\perp},\vec{r}_{2\perp}) \equiv
\frac{\langle\widetilde{\cal W}_L(\vec{l}_{1\parallel};\vec{r}_{1\perp};d)
\widetilde{\cal W}_L(\vec{l}_{2\parallel};\vec{r}_{2\perp};0)\rangle}
{\langle\widetilde{\cal W}_L(\vec{l}_{1\parallel};\vec{r}_{1\perp};d)\rangle
\langle\widetilde{\cal W}_L(\vec{l}_{2\parallel};\vec{r}_{2\perp};0)\rangle}
- 1,
\ee
where $d=(0,\vec{d}_{\perp},0)$, and, moreover,
\be
{\cal C}_L(\hat{l}_{1\parallel},\hat{l}_{2\parallel};\vec{d}_{\perp};
\vec{r}_{1\perp},\vec{r}_{2\perp}) \equiv \lim_{L_1,L_2\to\infty}
{\cal G}_L(\vec{l}_{1\parallel}, \vec{l}_{2\parallel};\vec{d}_{\perp};
\vec{r}_{1\perp},\vec{r}_{2\perp}),
\ee
where $L_i \equiv |\vec{l}_{i\parallel}|$ are defined to be the lengths of the
longitudinal sides of the loops in lattice units, and
$\hat{l}_{i\parallel} \equiv \vec{l}_{i\parallel}/L_i$. 
In the continuum limit, where $O(4)$ invariance is restored, we expect
\bea
\label{eq:contlim}
{\cal G}_L(\vec{l}_{1\parallel},\vec{l}_{2\parallel};\vec{d}_{\perp};
\vec{r}_{1\perp},\vec{r}_{2\perp}) &\mathop\simeq_{a\to 0}&
{\cal G}_E(\theta;T_1=aL_1/2,T_2=aL_2/2;a\vec{d}_{\perp};
a\vec{r}_{1\perp},a\vec{r}_{2\perp}), \nonumber \\
{\cal C}_L(\hat{l}_{1\parallel},\hat{l}_{2\parallel};\vec{d}_{\perp};
\vec{r}_{1\perp},\vec{r}_{2\perp}) &\mathop\simeq_{a\to 0}&
{\cal C}_E(\theta;a\vec{d}_{\perp};a\vec{r}_{1\perp},a\vec{r}_{2\perp}),
\eea
where $\hat{l}_{1\parallel}\cdot\hat{l}_{2\parallel} \equiv
\cos\theta$ defines the relative angle $\theta$ and $a$ is the lattice spacing.

\begin{figure}[htb]
\parbox{\halftext}{
\centerline{\includegraphics[width=6.4 cm] {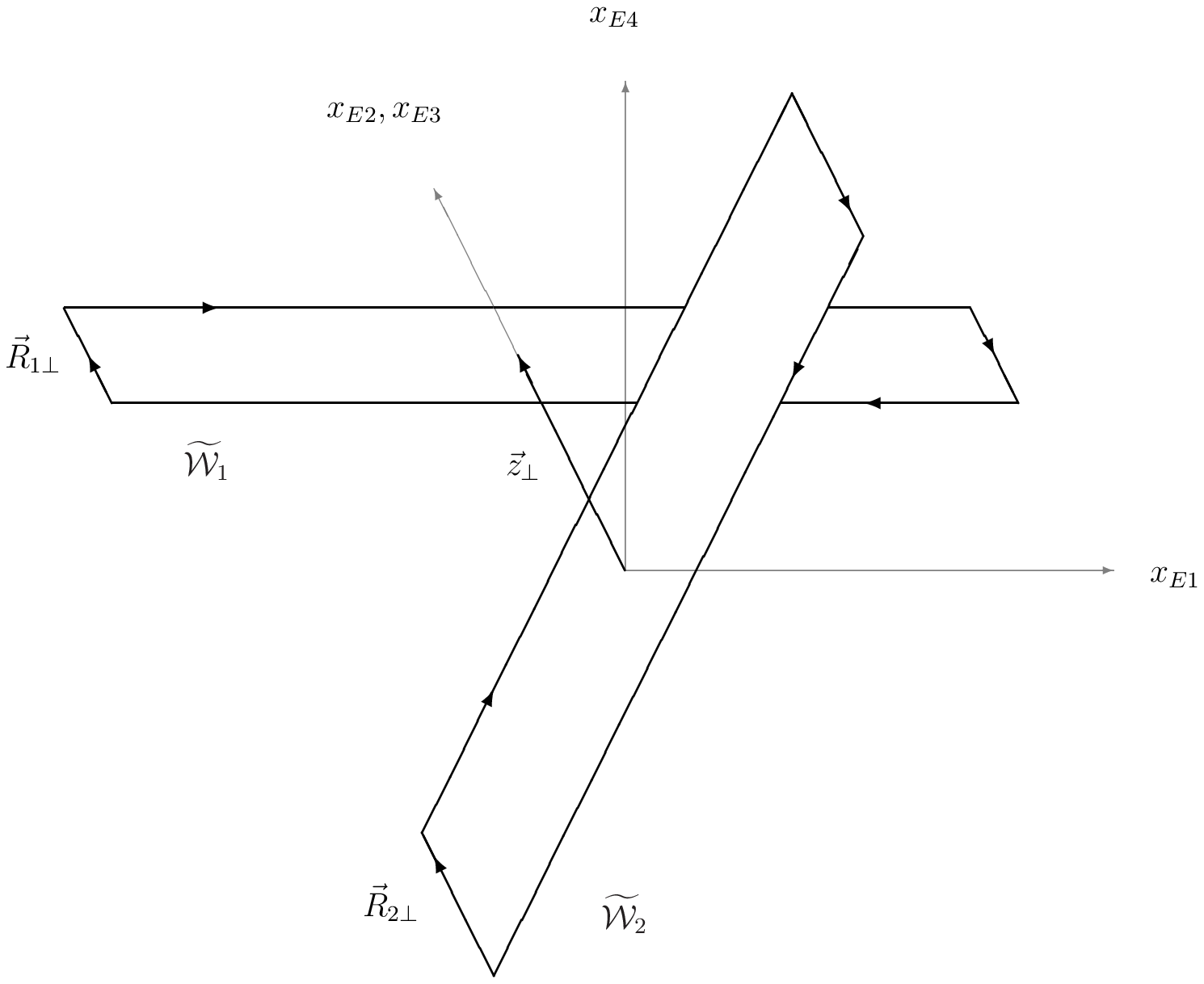}}
\caption{The relevant Wilson--loop configuration. Using the $O(4)$
invariance of the Euclidean theory we have put $p_{1E}$ parallel to
the $x_{E1}$ axis.}
\label{fig:loopconf}}
\hfill
\parbox{\halftext}{
\centerline{\includegraphics[width=6.4 cm] {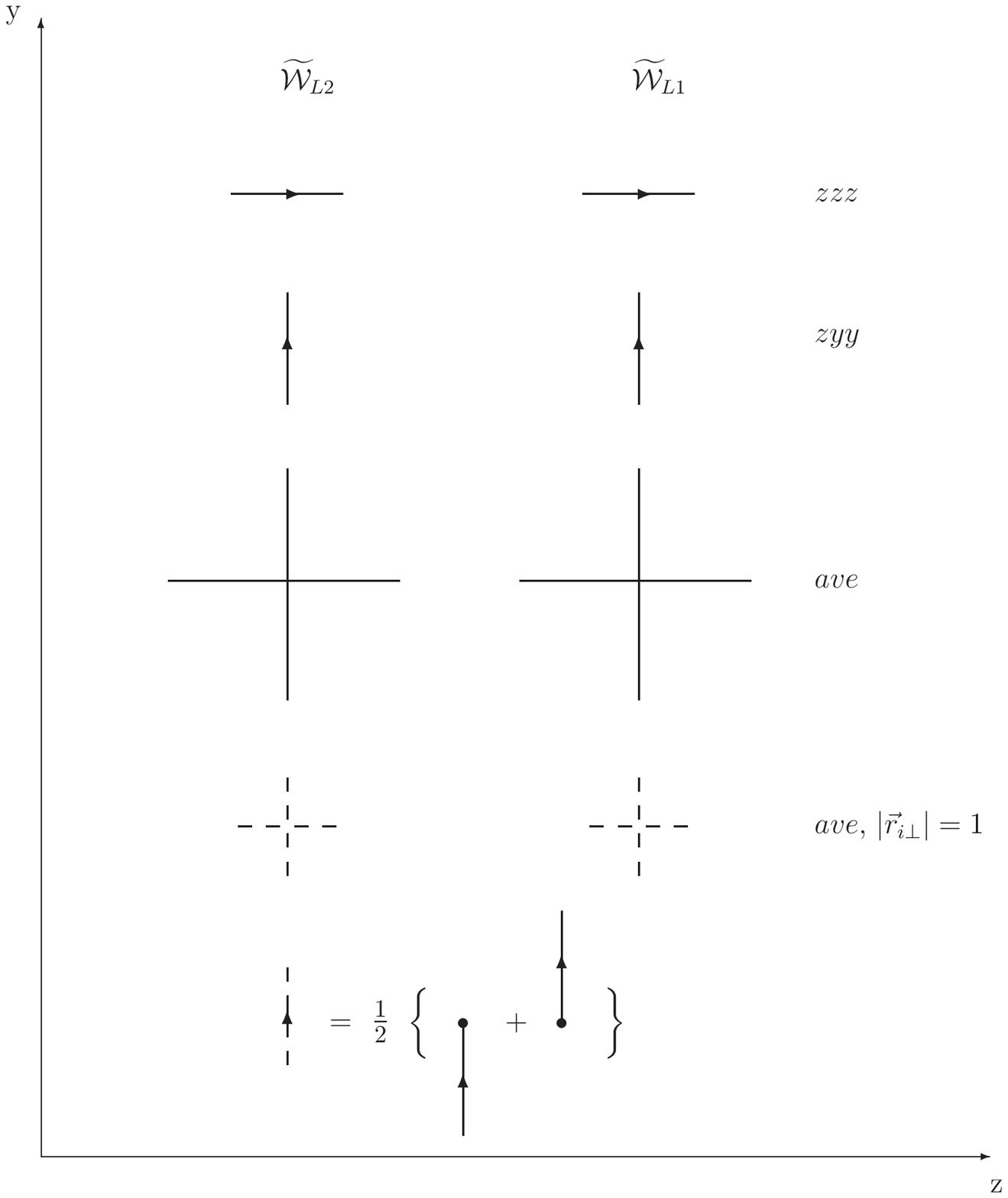}}
\caption{Loop configuration in the transverse plane. In the ``{\it ave}''
case the link orientation is not shown as it is averaged over.}
\label{fig:transvconf}}
\end{figure}

To keep the corrections due to $O(4)$ invariance breaking as small as
possible, we have kept one of the two loops {\it on--axis} and 
we have only tilted the other one as shown in Fig.~\ref{fig:loopconf};
the on--axis loop $\widetilde{\cal W}_{L1}$ is taken to be parallel
to the $x_{E1}$ axis, $\vec{l}_{1\parallel}=(L_1,0)$, and of length
$L_1=6,8$. We have used two sets of off--axis loops $\widetilde{\cal W}_{L2}$
tilted at $\cot \theta = 2,1,1/2,0,-1/2,-1,-2$, i.e., $\theta \simeq
26.565^{\circ}$, $45^{\circ}$, $63.435^\circ$, $90^\circ$, $116.565^\circ$,
$135^\circ$, $153.435^\circ$. We have used loops with transverse size
$|\vec{r}_{1\perp}|=|\vec{r}_{2\perp}|=1$ in lattice units; the loop
configurations in the transverse plane are those illustrated in
Fig.~\ref{fig:transvconf}, namely 
$\vec{d}_{\perp} \parallel \vec{r}_{1\perp} \parallel
\vec{r}_{2\perp}$ (which we call ``{\it zzz}'') and $\vec{d}_{\perp} \perp
\vec{r}_{1\perp} \parallel \vec{r}_{2\perp}$ (``{\it zyy}'').
We have also measured the orientation--averaged quantity
(``{\it ave}'') defined as
\be
\label{eq:ave}
{\cal C}_E^{ave}(\theta;\vec{z}_{\perp};|\vec{R}_{1\perp}|,|\vec{R}_{2\perp}|)
\equiv \int d\hat{R}_{1\perp} \int d\hat{R}_{2\perp}
{\cal C}_E(\theta;\vec{z}_{\perp};\vec{R}_{1\perp},\vec{R}_{2\perp}),
\ee
where $\int d\hat{R}_{i\perp}$ stands for integration over the orientations of
$\vec{R}_{i\perp}$. The lattice version of this equation is easily
recovered for even (integer) values of the transverse sizes; in our
particular case, $|\vec{r}_{i\perp}|=1$, we have to use a sort of
``{\it smearing}'' procedure, averaging nearby loops as depicted in
Fig.~\ref{fig:transvconf}.

\section{Numerical results and prospects}

In Refs.\cite{GM2008,GM2010}
we have performed a Monte Carlo calculation of the correlation function
${\cal G}_L$ of two Wilson loops for several values of the relative
angle, various lengths and different configurations in the
transverse plane, on a $16^4$ hypercubic lattice with periodic
boundary conditions. The link configurations were generated with the usual
Wilson action for $SU(3)$ {\it pure--gauge} theory, also
known in the literature as the {\it quenched} approximation of QCD, which
consists in neglecting dynamical fermion loops by setting the fermion
matrix determinant to a constant.

We have measured the correlation functions $\langle \widetilde{\cal W}_{L1}
\widetilde{\cal W}_{L2} \rangle$ and the loop expectation
values $\langle \widetilde{\cal W}_{L1}\rangle$ and $\langle
\widetilde{\cal W}_{L2} \rangle$, with \mbox{$\widetilde{\cal W}_{L1} \equiv
\widetilde{\cal W}_{L}(\vec{l}_{1\parallel};\vec{r}_{1\perp};d)$} and
\mbox{$\widetilde{\cal W}_{L2} \equiv
\widetilde{\cal W}_{L}(\vec{l}_{2\parallel};\vec{r}_{2\perp};0)$}, on 30000
thermalised configurations at $\beta\equiv 6/g^2=6.0$.
As it is well known, the lattice spacing $a$ is related to the bare coupling
constant $g$ (i.e., to $\beta$) through the renormalisation group
equation. The lattice scale, i.e., the value of $a$ in physical units, is
determined from the physical value of some relevant (dimensionful) observable
like the string tension or the static $q\bar{q}$ force at some fixed distance:
in our case one finds that $a(\beta=6.0)\simeq 0.1\,{\rm fm}$.
The choice of $\beta=6.0$ on a $16^4$ lattice is made in
order to stay within the so--called ``{\it scaling window}'': in this sense we
are relying in an indirect way on the validity of the relation
(\ref{eq:contlim}) between Wilson--loop correlation functions on 
the lattice and in the continuum (and therefore we shall use the notation
${\cal G}_E$/${\cal C}_E$ of the continuum in all the figures reporting our
lattice data).

\begin{figure}[htb]
\parbox{\halftext}{
\centerline{\includegraphics[width=6.4 cm] {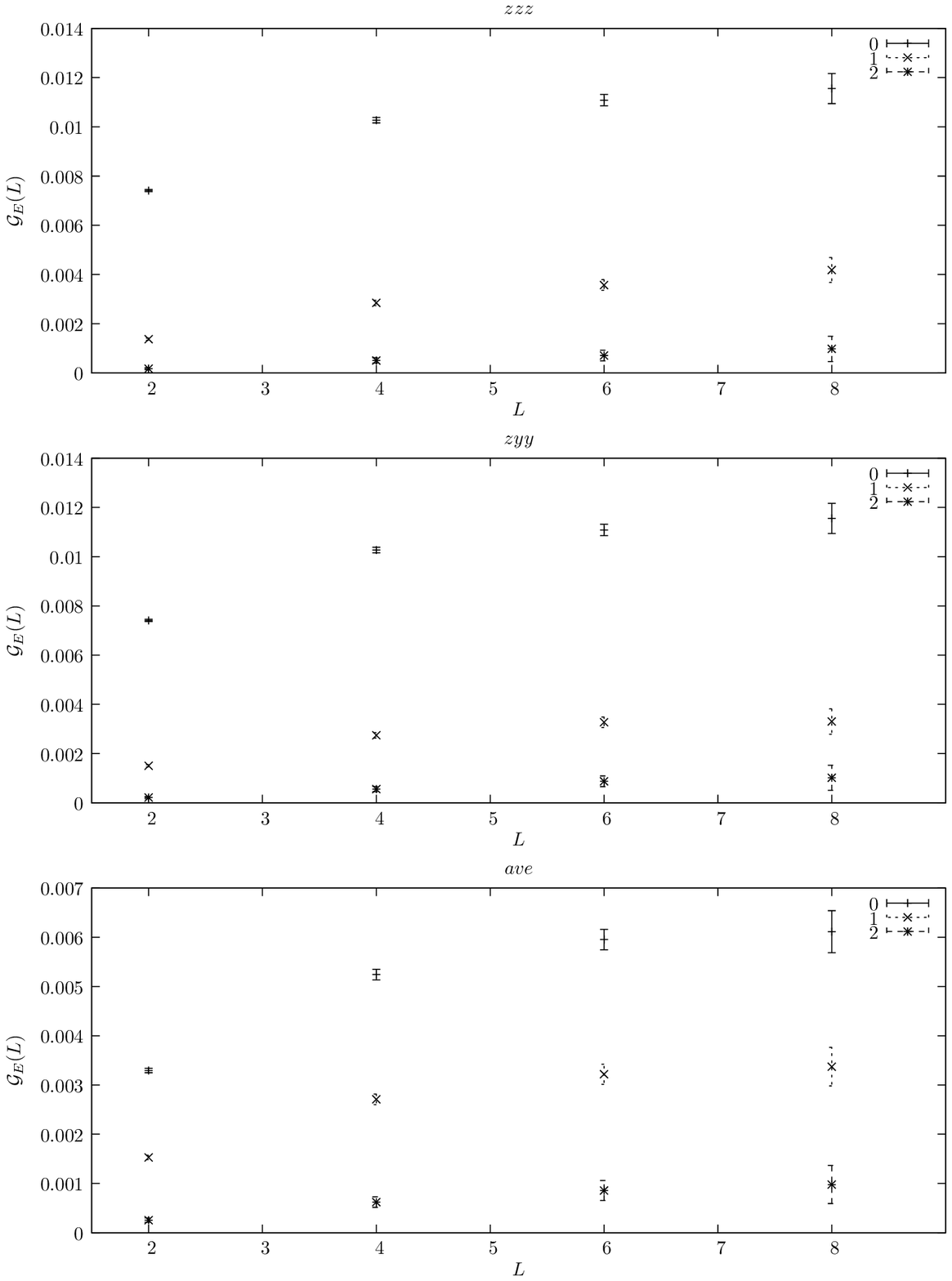}}
\caption{Dependence of ${\cal G}_E$ on the length $L_1=L_2=L$ (in lattice
units) of the loops at $\theta=90^{\circ}$ for $d=0,1,2$.}
\label{fig:Tdep}}
\hfill
\parbox{\halftext}{
\centerline{\includegraphics[width=6.4 cm] {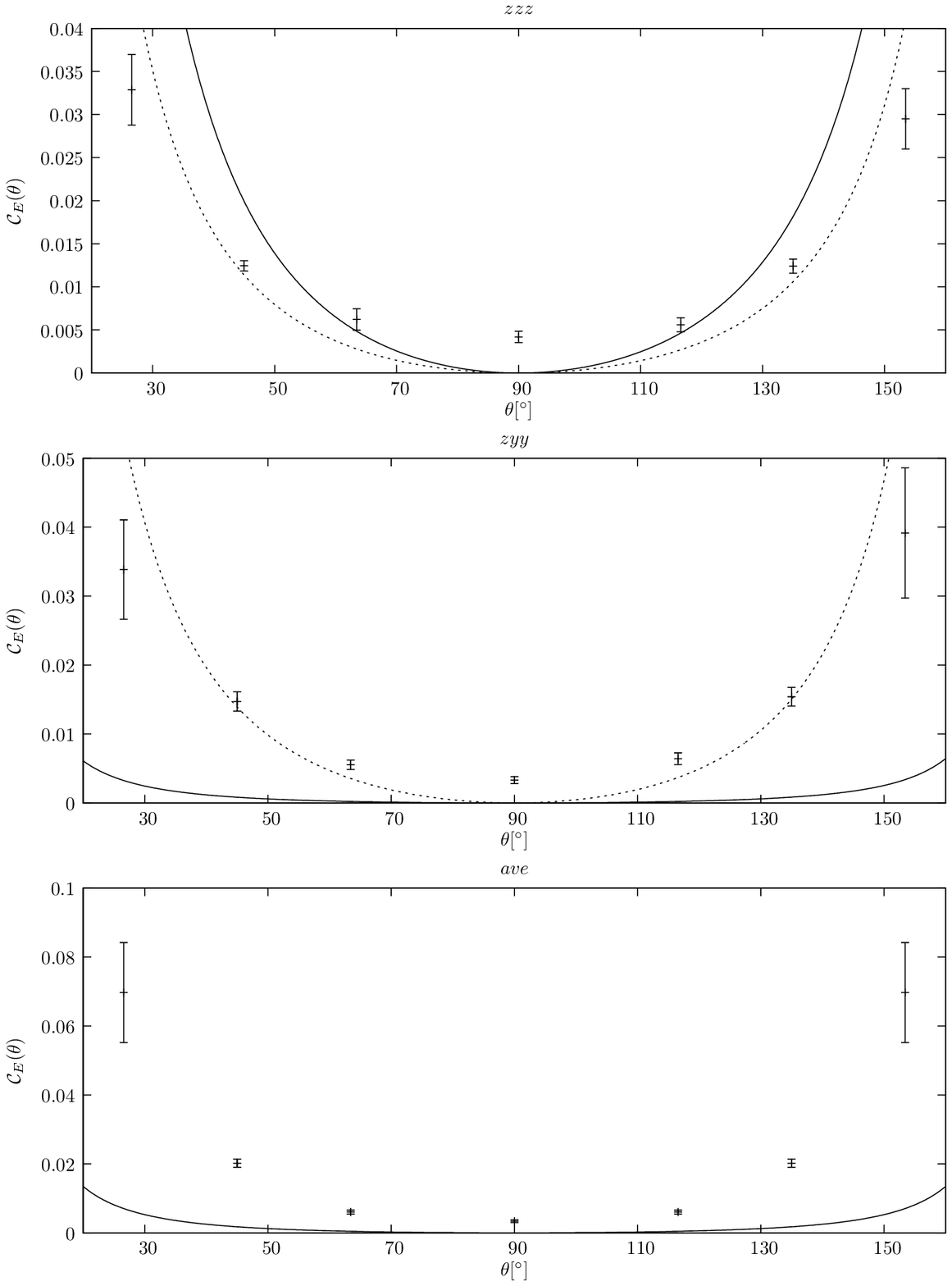}}
\caption{Comparison of the lattice data to the SVM prediction (\ref{eq:SVM})
with $K_{\rm SVM}$ calculated according to Ref.~\cite{LLCM2} (solid line) 
and to the one--parameter ($K_{\rm SVM}$) best--fit (for the ``{\it zzz}''
and ``{\it zyy}'' cases only) with the SVM expression (\ref{eq:SVM}) 
(dotted line) at $d=1$.}
\label{fig:svm1}}
\end{figure}

As explained in Section 2, we are interested in the $T\to\infty$
limit and so we have to somehow perform it on the lattice. In
practice, we have to look for a {\it plateau} of the correlation
function plotted against the loop lengths $L_1$ and $L_2$: in
Fig.~\ref{fig:Tdep} we show the dependence of the correlator on the length 
$L_1=L_2=L$ of the loops at $\theta=90^{\circ}$. Of course, on
a $16^4$ lattice it is difficult to have a sufficiently long loop 
while at the same time avoiding finite size effects and at best
we can push the calculation up to $L=8$; nevertheless, a {\it plateau} seems 
to have been practically reached at about $L=L_{\rm pl} \simeq
6\div 8$. As $\theta$ varies from $90^{\circ}$ towards $0^{\circ}$ or
$180^{\circ}$, we expect $L_{\rm pl}$ to grow. Indeed, $L_{\rm pl}$ blows up at 
$0^{\circ},180^{\circ}$ due to the relation between the correlation function
and the static dipole--dipole potential $V_{dd}$ (see Refs.~\cite{GM2008,GM2010}
and references therein: some preliminary lattice data for $V_{dd}$ have been
obtained in Ref.~\cite{GM2010}):
\be
\label{eq:Pot}
{\cal G}_E(\theta = 0;T;\vec{z}_\perp;\vec{R}_{1\perp},\vec{R}_{2\perp})
\mathop\simeq_{T \to \infty} \exp \left[
-2T~ V_{dd}(\vec{z}_\perp,\vec{R}_{1\perp},\vec{R}_{2\perp}) \right] -1,
\ee
from which we expect ${\cal G}_E$ to diverge at $\theta=0^{\circ}$
and, by virtue of the {\it crossing--symmetry relations},\cite{GM2006}
also at $\theta=180^{\circ}$.
In the following we will consider only $\theta\neq 0^{\circ},180^{\circ}$:
our data show that the correlation
function is already quite stable against variations of the loop lengths at
$L_1,L_2\simeq 8$ (at least for $\theta$ not too close to $0^{\circ}$ or
$180^{\circ}$) and so we can take the data for the largest loops available as a 
reasonable approximation of ${\cal C}_L$, defined as the asymptotic value of
${\cal G}_L$ as $L_1,L_2\to\infty$.

We have considered the values $d=0,1,2$ for the distance between the
centers of the loops: as expected, the correlation functions vanish rapidly
as $d$ increases, thus making the calculation with our simple ``brute force''
approach very difficult at larger distances.

From now on we will discuss the issue of the angular dependence of the
correlation function. 
As already pointed out in the Introduction, numerical simulations of
LGT can provide the Euclidean correlation function only for a finite
set of $\theta$--values, and so its analytic properties cannot be
directly attained; nevertheless, they are first--principles
calculations that give us (inside the errors) the true QCD expectation
for this quantity. Approximate analytic calculations of this same
function then have to be compared with the lattice 
data, in order to test the goodness of the approximations involved. 
The Euclidean correlation functions we are interested in have been
evaluated in the {\it Stochastic Vacuum Model} (SVM),\cite{LLCM2}
in the {\it Instanton Liquid Model} (ILM)~\cite{instanton1,GM2010}, and
using the AdS/CFT correspondence:\cite{JP1} the comparison of our data with
these analytic calculations is not, generally speaking, fully satisfactory.

In the SVM~\cite{LLCM2} the Wilson--loop correlation function
is given by the expression
\be
\label{eq:SVM}
{\cal C}^{\,\rm (SVM)}_E(\theta) = \frac{2}{3}
\exp\left(-\frac{1}{3} K_{\rm SVM}\cot\theta\right)
+ \frac{1}{3} \exp\left(\frac{2}{3} K_{\rm SVM}\cot\theta\right) - 1,
\ee
where $K_{\rm SVM}$ is a function of $\vec{z}_{\perp}$, $\vec{R}_{1\perp}$
and $\vec{R}_{2\perp}$ only, whose precise expression, given in
Ref.\cite{LLCM2}, we have used to numerically evaluate the correlator
(\ref{eq:SVM}) in the relevant cases.
The SVM prediction (\ref{eq:SVM}) agrees with our lattice data in a few
cases, at least in the shape and in the order of magnitude, but, in general,
it is far from being satisfactory: for example, the comparison
with our data for $d=1$ is shown in Fig.~\ref{fig:svm1}.
More or less the same conclusion is reached if one instead performs a
one--parameter ($K_{\rm SVM}$) best--fit with the given expression:
the values of the chi--squared per degree of freedom ($\chi^2_{\rm d.o.f.}$)
of this and the other fits that we have performed are listed in
Table \ref{tab:chi2}.

\begin{table}[t]
\caption{Chi--squared per degree of freedom for a best--fit with the
indicated function.}
\centering
\begin{tabular}[h]{l|cc|ccc|ccc}
\hline
\hline
$\chi^2_{\rm d.o.f.}$ & \multicolumn{2}{c|}{$d=0$} &
\multicolumn{3}{c|}{$d=1$} & \multicolumn{3}{c}{$d=2$}\\
& {\it zzz}/{\it zyy} & {\it ave} & {\it zzz} & {\it zyy} &
{\it ave} & {\it zzz} & {\it zyy} & {\it ave} \\
\hline
SVM & 51 & - & 16 & 12 & - & 1.5 & 2.2 & -\\
pert & 53 & 34 & 16 & 13 & 13 & 1.5 & 2.2 & 4.5\\
ILM & 114 & 94 & 14 & 15 & 45 & 0.45 & 0.35 & 1.45\\
ILMp & 20 & 9.4 & 0.54 & 0.92 & 1.8 & 0.13 & 0.12 & 0.19\\
AdS/CFT & 40 & - & 1 & 0.63 & - & 0.14 & 0.065 & -\\
\hline
\end{tabular}
\label{tab:chi2}
\end{table}

We have also tried best--fits with the following simple functional forms:
\bea
\label{eq:pert}
{\cal C}_E^{(\rm pert)}(\theta) &=& K_{\rm pert} (\cot\theta)^2,\\
\label{eq:instanton}
{\cal C}^{\,\rm (ILM)}_E(\theta) &=& \frac{K_{\rm ILM}}{\sin\theta}.
\eea
The first expression \eqref{eq:pert} is exactly what one obtains in
leading--order perturbation theory.\cite{BB,Meggiolaro05,LLCM2}
The second expression \eqref{eq:instanton} is the one predicted by
one--instanton effects in the ILM.\cite{instanton1,GM2010}
The results, shown in Table \ref{tab:chi2}, are again not satisfactory.
In particular, the ILM expression seems to be strongly disfavoured at $d=0$,
while at $d=2$ it looks better than the SVM and perturbative--like expressions.

\begin{figure}[htb]
\parbox{\halftext}{
\centerline{\includegraphics[width=6.4 cm] {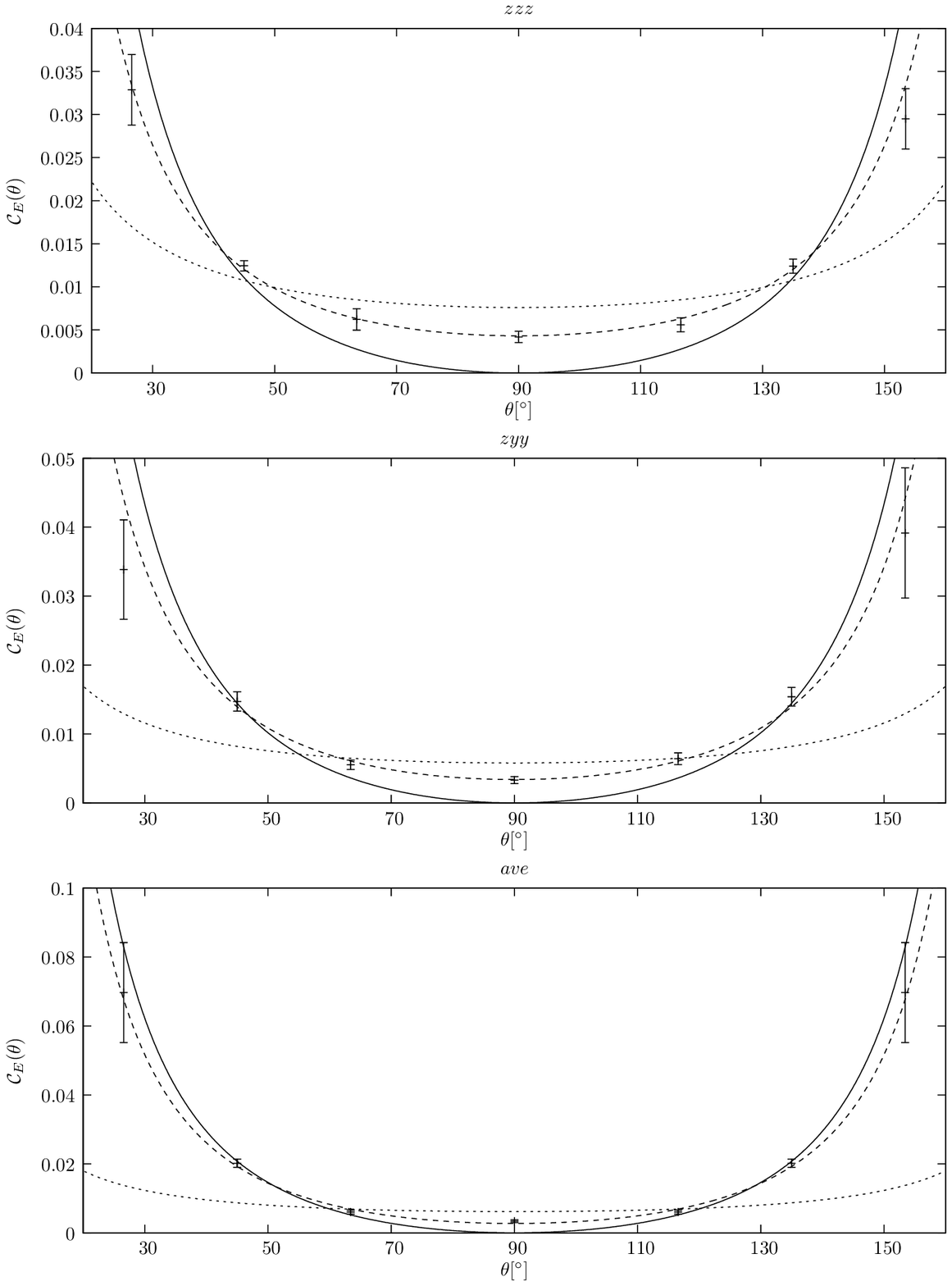}}
\caption{Comparison of lattice data to best--fits with the
perturbative--like expression (\ref{eq:pert}) (solid line), the
ILM expression (\ref{eq:instanton}) (dotted line) and the ILMp
expression (\ref{eq:instpert}) (dashed line) at $d=1$.}
\label{fig:pertinst1}}
\hfill
\parbox{\halftext}{
\centerline{\includegraphics[width=6.4 cm] {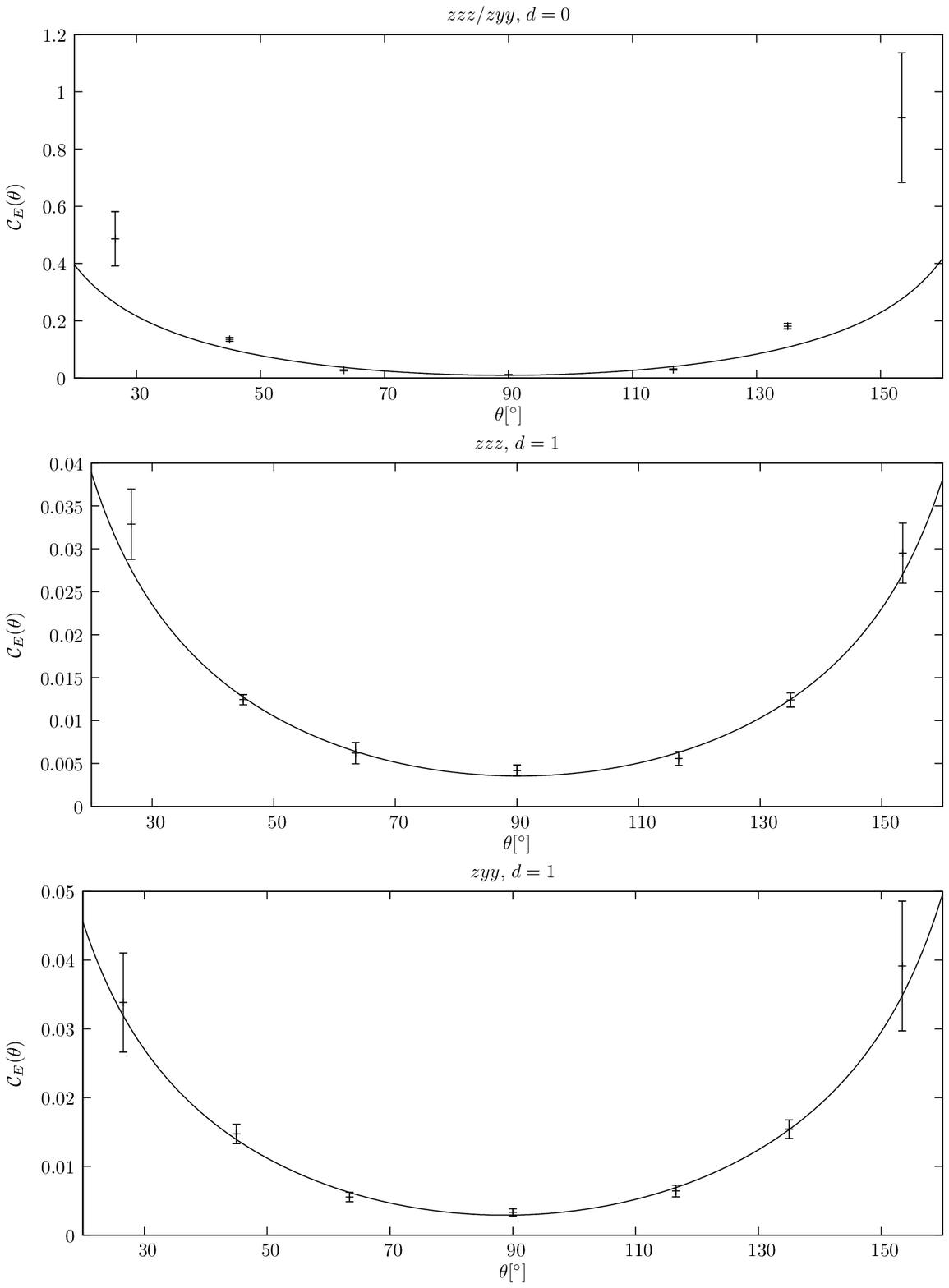}}
\caption{Comparison of lattice data to a best--fit with the AdS/CFT
expression (\ref{eq:AdS}) for various cases.}
\label{fig:ads}}
\end{figure}

By combining the two previous expressions into the following expression,
\be
\label{eq:instpert}
{\cal C}^{\,\rm (ILMp)}_E(\theta) = \frac{K_{\rm ILMp}}{\sin\theta}
+ K_{\rm ILMp}'(\cot\theta)^2,
\ee
largely improved best--fits have been obtained, as one can see in
Table \ref{tab:chi2}. The resulting best--fit functions in the $d=1$
cases are plotted in Fig.~\ref{fig:pertinst1}.

A well--defined numerical {\it prediction} for the prefactor $K_{\rm ILM}$ of
$1/\sin\theta$ in the ILM has been obtained in Ref.:\cite{GM2010}
in Table \ref{tab:insttab} we compare this prediction with the value obtained
with a fit to the lattice data with the fitting functions \eqref{eq:instanton}
and \eqref{eq:instpert}. The ILM prediction turns out to be more or less of the
correct order of magnitude in the range of distances considered, at least
around $\theta=\pi/2$, but it does not match properly the lattice data.
The agreement with the data seems to be quite good at $d=2$;
however, concerning the dependence on the relative distance between the loops,
it seems that the ILM overestimates the correlation length which sets the scale
for the rapid decrease of the correlation function. This is also supported by
the comparison of the instanton--induced {dipole--dipole potential} $V_{dd}$
with some preliminary numerical results on the lattice.\cite{GM2010}

\begin{table}[t]
\caption{Value of $K_{\rm ILM}\times 10^3$ for the relevant configurations: 
ILM prediction (first column), ``ILM'' fit (second column) and ``ILMp'' fit
(third column).}
\centering
\begin{tabular}{l|cc|cc|cc}
\hline
\hline
& \multicolumn{2}{c|}{predicted} &
\multicolumn{2}{c|}{fitted--ILM} &
\multicolumn{2}{c}{fitted--ILMp}\\
$d$ & $zzz$ & $zyy$ & $zzz$ & $zyy$ & $zzz$ & $zyy$ \\
\hline
0 &0.880--1.08 & 0.880--1.08 &17.1 & 17.1 & 9.86 & 9.86 \\
1 &0.827--1.02 & 0.798--0.984 & 7.60 & 5.79 & 4.32 & 3.39 \\
2 &0.692--0.853 & 0.607--0.748 &1.31 & 1.43 &0.947 & 1.14 \\
\hline
\end{tabular}
\label{tab:insttab}
\end{table}

Finally, we have tried a best--fit with the expression that one obtains
using the AdS/CFT correspondence, for the ${\cal N}=4$ SYM theory at large
$N_c$, large 't Hooft coupling and large distances between the loops:\cite{JP1}
\be
\label{eq:AdS}
{\cal C}^{\,\rm (AdS/CFT)}_E(\theta)
= \exp\left\{\dfrac{K_1}{\sin\theta} + K_2\cot\theta +
K_3\cos\theta\cot\theta\right\}-1.
\ee 
The results are shown in Table \ref{tab:chi2} and in Fig.~\ref{fig:ads}.
Taking into account that this is a three--parameter best--fit, even this one
is not satisfactory: best--fits with QCD--inspired
expressions with only two parameters, like, e.g., the ILMp expression
(\ref{eq:instpert}) [or some appropriate modification of the SVM expression
(\ref{eq:SVM})] give smaller $\chi^2_{\rm d.o.f.}$.

As we have said in the Introduction, the main motivation in studying soft
high--energy scattering is that it can lead to a resolution of the total cross
section puzzle, so it is worth discussing what the various models have to say
on this point. Using Eqs.~\eqref{scatt-loop}, \eqref{analytic_C} and the
{\it optical theorem}, it is easy to see that the SVM and the ILMp expressions
give {\it constant} cross sections at high energy,
as in these cases the high--energy limit can be carried over under the
integral sign, so that the knowledge of the $\theta$--dependence of
the correlation function is sufficient to completely determine,
after the analytic continuation $\theta \to -i\log(s/m^2)$,
the high--energy behaviour of total cross sections.
Although the AdS/CFT expression (\ref{eq:AdS}) is not, of course,
expected to describe real QCD, it nevertheless shows how a
non--trivial high--energy behaviour could emerge from a simple
analytic dependence on the angle $\theta$. In this case,
after the analytic continuation into Minkowski space--time, it is not
possible to pass to the high--energy limit under the integral sign, as
the integrand is an oscillating function of the energy, and one should
carry over the remaining integrals first.
The experimentally observed universality in the high--energy behaviour suggests
that the integration over the distance between the loops should be the
relevant one: this seems to be the case also in the AdS/CFT case, where,
combining the knowledge of the various coefficient functions in
\eqref{eq:AdS} in the large impact--parameter region~\cite{JP1} with the
unitarity constraint in the small impact--parameter region,
a variety of possible high--energy behaviours for the total cross
section is shown to emerge (including, e.g., a {\it pomeron}--like behaviour
$\sigma \sim s^{1/3}$).\cite{GP2010}\\
It seems then worth investigating further the dependence of the
correlation functions on the relative distance between the loops, as
well as on the dipole sizes, as they could combine in a
non--trivial way with the dependence on the relative angle:
these and other related issues will be addressed in future works.

As a final and important remark, we note that our data show a clear signal of
$C$--odd contributions in dipole--dipole scattering.
Because of the {\it crossing--symmetry relations} \eqref{crossing_C},
it is natural to decompose the Euclidean correlation function
${\cal C}_E(\theta)$ as a sum of a {\it crossing--symmetric} function
${\cal C}_E^+(\theta)$ and a {\it crossing--antisymmetric} function
${\cal C}_E^-(\theta)$, with ${\cal C}_E^{\pm}(\theta) \equiv \frac{1}{2}
\left[ {\cal C}_E(\theta) \pm {\cal C}_E(\pi -\theta) \right]$.
\cite{Meggiolaro07}
Upon analytic continuation from the Euclidean to the
Minkowskian theory and using Eq.~\eqref{crossing_C}, one can show that
they are related respectively to {\it pomeron} (i.e., $C=+1$) and
{\it odderon} (i.e., $C=-1$) exchanges in the dipole--dipole
scattering amplitude.
Looking at our lattice data, we notice that there is an asymmetry with respect
to $\theta=\pi/2$ in the plot of the Euclidean correlation function, for the
``$zzz$'' and ``$zyy$'' tranverse configurations, against the relative angle.
(As regards the orientation--averaged quantity ${\cal C}_E^{ave}$,
defined in Eq. \eqref{eq:ave}, it is trivially {\it crossing--symmetric}
by virtue of Eqs. \eqref{crossing_C}.)
In other words, a small but non--zero {\it crossing--antisymmetric}
component ${\cal C}_E^-$ is present in our data, thus
signalling the presence of {\it odderon} contributions to 
the loop--loop correlation functions and in turn to the dipole--dipole
scattering amplitudes.
Even though these $C$--odd contributions are averaged to zero
in meson--meson scattering (at least in our model, as long as
the squared meson wave functions satisfy some reasonable symmetry
properties in their dependence on the dipole orientations and on the
longitudinal--momentum fractions), they might play a non--trivial role
in more general hadron--hadron processes in which baryons and
antibaryons are also involved.

\section*{Acknowledgements}
E.M. thanks the Yukawa Institute for Theoretical Physics of Kyoto University
for having funded his stay during the Symposium of the YIPQS
International Workshop ``High Energy Strong Interactions 2010''.
He also wants to thank Dr. Kazunori Itakura for the invitation to the
Symposium and for useful discussions.

\end{document}